\documentclass[preprint,12pt]{elsarticle}
\usepackage[utf8]{inputenc}
\usepackage{makecell}
\usepackage{diagbox}
\usepackage{amsmath}
\usepackage{pdflscape}
\usepackage{natbib}

\usepackage{graphicx}
\usepackage{adjustbox}
\usepackage{lipsum}
\usepackage{float}

\journal{Physica A}
\begin{document}

\begin{frontmatter}

\title{String percolation threshold for elliptically bounded systems}


\author{J. E. Ram\'irez$^{a}$\footnote{E-mail address:  jerc.fis@gmail.com}, A. Fern\'andez T\'ellez$^a$, I. Bautista$^{a,b}$\footnote{E-mail address:  irais.bautista@fcfm.buap.mx} \\
\small $^{a}$ Facultad de Ciencias F\'isico Matem\'aticas, Benem\'erita Universidad Aut\'onoma de Puebla, 1152, M\'exico.\\
\small $^b$C\'atedra CONACyT, CONACyT, 03940 Ciudad de M\'exico, M\'exico}

\address{}

\begin{abstract}
It has been shown that a hot and dense deconfined nuclear matter state produced in ultra-relativistic heavy-ion collisions, can be quantitatively described by the String Percolation phenomenological model. The model address the phase transition in terms of the two-dimensional continuum percolation theory over strings, which are schematic representations of the fundamental interactions among the partons of the colliding nuclei in the initial state.  In this work, we present an extension of the critical string density results including the eccentricity dependence on the initial state geometry focus on small string number with different density profile, small deviations from the different profile densities are found. The percolation threshold shows consistency with the thermodynamic limit for the uniform density profile with a large number of strings in the case of circular boundary system. A significant dependence on the eccentricity for a small number of strings compared to high occupancy systems is exhibited, the implications may become relevant in hadron-hadron or hadron-nucleus collision systems.
 
\end{abstract}


\end{frontmatter}



\section{Introduction}
It has been six decades from the introduction of the percolation problem in statistical physics \cite{Broadbent:1957rm}, which has derived in several applications to a wide number of problems in statistical physics with relevance to many fields \cite{Hoshen:1976zz}, going from the description of the formation of galactic structures \cite{schulman1986,Klypin:1993rz} to the study of the most fundamental components of the nuclear matter 
\cite{Biro:1986ebt,Baym,Campi:1986nxv}.

Generally, the percolation theory is closely associated with the study of phase transitions and transport phenomena \cite{Broadbent:1957rm,Kirkpatrick,Stauffer:1978kr,Elliott:2002dk}. In this context, the transport phenomena occur if a spanning cluster emerges for a given density of objects in such case, the process it says that  percolate (or crossovers the system) \cite{Stauffer:1978kr,Stauffer-book,Saberi}. The determination of the critical density of objects, namely percolation threshold in a given system (conditions needed to percolate) is one of the goals of the subject \cite{Saberi,feng}.

The percolation probability $P$ is commonly defined as the fraction of occupied/connected sites belonging to the spanning cluster \cite{Stauffer:1978kr,Stauffer-book}. In the infinite size limit, a second order phase transition is shown around the percolation threshold. The order parameter $P$ follows the power law $P\propto (p-p_c)^\beta$, for the occupied/connected probability ($p$) larger or equal to the percolation threshold ($p_c$) with $\beta$ being the critical exponent. In particular $\beta=5/36$ in two-dimensions
 \cite{Stauffer:1978kr,Saberi,sykes,Essam}.

In the following manuscript we will use the framework of the String Percolation Model (SPM) which applies a two-dimensional continuum percolation theory to the high energy and temperature finite system, formed in nuclear and hadron collisions, where the Quantum Chromodynamic theory (QCD), predicts a phase transition to a quark-gluon deconfined state called Quark Gluon Plasma (QGP) phase, that was formed at the first microseconds of the early Universe \cite{Rafelski:2013obw}. Probes of the QGP phase had been found by large experimental efforts at the Relativistic Heavy Ion Collider (RHIC) and the Large Hadron Collider (LHC) \cite{Bala:2016hlf,Pasechnik:2016wkt,Armesto:2015ioy,Foka:2016vta}.

In relativistic heavy-ions collisions, due to Lorentz contraction, the two interacting hadron/nucleus can be seen as two discs which interaction gives rise to an overlap area $S$, which can be projected on the impact parameter plane. In the SPM to consider the color interaction among the partons of the projectiles, we will throw randomly $N$ penetrable discs (namely strings) coming from the projection on the impact parameter plane of extended objects (color flux tubes) along the rapidity axis with color charges at their ends. The color of these strings is confined in a projected area, which in the impact parameter transverse plane is considered as a small area $S_{0} = \pi r_{0}^{2}$, with $r_{0}\sim.25fm$ \cite{Braun:2015eoa,Bali:1994de,Schlichter:1994dc}.

The transverse nuclear overlap area is consider as a two dimensional flat surface $S$, usually represented by a circle of radius $R$, and the transverse strings of radius $r_0$, which are randomly placed into $S$ \cite{Nardi:1998qb,Amelin:1993cs}. The formed discs are allow to overlap. This last condition allows the system to form clusters of strings whose geometrical pattern is governed by the percolation theory \cite{Braun:2015eoa,Braun:2001us}. The number $N$ of strings in the system depends on the energy and atomic number of the colliding nucleus \cite{Amelin:1993cs}. In the SPM a consistent description of the multiplicity production in AA collisions has been found, with a good description of strongly interacting matter on both sides of the percolation threshold whose results agree rather well with other models \cite{Celik:1980td}. 

Signatures of collective effects in pp and proton lead (p-Pb) collisions at the LHC and RHIC energies are consistent with the formation of a string density comparable to the one created in nucleus-nucleus (AA) collisions \cite{Bautista:2015kwa,I.Bautista:2017}. The obtained string density for each multiplicity class in pp collision at 7 TeV from the mean transverse momentum distribution can relate the number of strings, by the string density definition \cite{Bautista:2015kwa}.

The variation of the initial string number can be considered as a direct correspondence by varying the fixed area instead of varying the density itself \cite{Bautista:2015kwa}.

But so far to apply these results to pp collision systems we may consider further differences that  become relevant in small collision systems. Since we are treating with one of the lowest dense system in terms of valence quarks one should consider that the main contribution to the particle production for higher multiplicities is due to the contribution of the sea quarks and gluon exchange of the colliding protons. Nevertheless, since the system is confined to a small area, higher string density can still be produced with enough high energy density. One also has to consider that since the area is small, the modification of the initial geometry should become more relevant that in the case of AA collisions. The importance of the initial state geometry in such systems including the elliptic geometry and its eccentricity have been recently studied in references \cite{Schenke:2014jza, Gelis:2016upa}.

It is known that the description of the QGP phase transition has a significant dependence on the initial state geometry, which is considered the origin of the observed azimuthal asymmetry \cite{Bautista:2009my}. In consistency with this fact, we implement the eccentricity of the elliptical boundary as a first initial state geometry effect in string percolating systems. Further initial geometries can be considered, but in this work, we will limit ourselves to the geometry related to the second Fourier coefficient of the flow the so-called elliptic flow which reflects the event-by-event eccentricity.

This paper is organized as follows: In Section 2, we describe the model of a continuum percolation system with finite number of strings and the elliptical boundary conditions. In Section 3, we show the results obtained for the percolation probability and the percolation threshold as function on eccentricity, the number of strings, for Uniform and Gaussian density profiles. Also, the results are compared with the thermodynamic limit considered by around 500 strings in the Uniform density profile system \cite{Rodrigues:1998it}. Finally, we present our conclusion in Section 4.

\section{A Monte Carlo model for string percolating systems}

The continuum percolation is defined on a continuous flat surface  with  the  string density  $n$, defined as the quotient between the number of objects on $S$ and the area $S$, i.e., $n=N/S$ as the main parameter to  describe the system. Equivalently, the filling factor is defined as $\eta=an$, where $a$ is the area of one object in the system \cite{Mertens}. As usual on continuum percolation studies, we will describe the string percolating system through the filling factor.

The percolation threshold can be determined using Monte Carlo methods \cite{Reynolds:1980zz,Ziff,Newman}. Thus, we implement a kind of find-union algorithm to simulate the string percolating systems.

We will start by considering a fixed number $N$ of strings with radius $r_0$, randomly distributed on an ellipse determined by the major and minor semi-axes $a$ and $b$, respectively, without periodic boundary conditions. The string population fills the ellipse area with a filling factor $\eta$ defined by: 
\begin{equation}
\eta=\frac{r_0^2N}{ab}.
\end{equation}
Also, $a$ and $b$ are related each other by the eccentricity $\varepsilon$ \cite{wexler}:
\begin{equation}
\varepsilon =\sqrt{1-\frac{b^2}{a^2}}.
\end{equation}
Thus, for a given $\eta$, $\varepsilon$, $N$, we can compute the major and minor semi-axes for the elliptical bounded of the string percolating system by the following equations
\begin{eqnarray}
a^2=\frac{r_0^2N}{\eta \sqrt{1-\varepsilon^2}};& &b^2=\frac{r_0^2N\sqrt{1-\varepsilon^2}}{\eta}.
\end{eqnarray}
Once determined $a$ and $b$, we take a random point $(x,y)$ distributed in the rectangle $[-(a-r_0),a-r_0]\times[-(b-r_0),b-r_0]$ according with a density profile. This point is the center of the string and it is included in the string population if satisfies the following condition
\begin{equation}
\frac{x^2}{(a-r_0)^2}+\frac{y^2}{(b-r_0)^2}\leq 1.
\end{equation}
The constrictions will only consider strings completely embedded in the elliptical region.
In this way, we can generate all $N$ strings needed to build the percolating system.
Note that with this construction, in the limit $\varepsilon=0$, we recover the particular case for a SPM bounded by circles, which has been already studied by several authors \cite{Amelin:1993cs,Braun:2001us,Celik:1980td,Satz:2002ku}.

\begin{figure}[ht]
\centering
\includegraphics[scale=1]{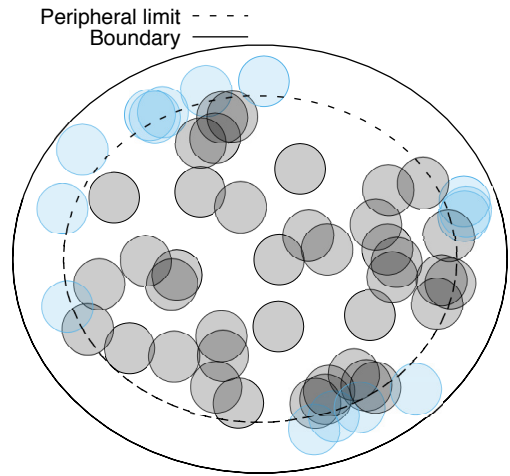}
\caption{Scheme percolation of Uniformly distributed strings, with parameters $N=55$, $\eta=0.7$, $\varepsilon=0.50$. The solid line is the elliptic boundary of the system and the dashed line represents the internal peripheral limit needed to define a spanning cluster. Blue circles are the peripheral strings satisfying the relation in Eq.~\eqref{per}.}
\label{fig-perif}
\end{figure}

After building the string percolating system with exact $N$ strings of radius $r_0$, randomly distributed on the ellipse, to form the clusters in the system we consider two discs being the nearest neighbors if $|r_i-r_j|\leq r_0$, with $r_i,r_j$ being the position vectors of two objects labeled $i,j$, respectively.
In this way, we assure that a cluster will become a set of mutually overlapped discs.

Nevertheless, to establish the emergence of the spanning cluster we need one more requirement: We define a peripheral discs as the one satisfying the following condition: 
\begin{equation}
\frac{x^2}{(a-2r_0)^2}+\frac{y^2}{(b-2r_0)^2}>1.
\label{per}
\end{equation}
In Fig.\ref{fig-perif}, we show an example of a string system with the peripheral condition given by Eq.~\eqref{per}. The peripheral strings (blue discs) lie in the region between the elliptical boundary (solid line) and the peripheral limit (dashed line).
In this way, we assure that there is a spanning cluster in the string system if the largest cluster has more than one peripheral string and the largest distance between the peripheral strings is greater than $2(b-2r_0)$. This condition is imposed in order to ensure that the spanning cluster at least cross-over the system through the minor semi-axes.
Since we have determined whether there is a spanning cluster for the generated system, we calculate the percolation probability as the rate between the number of strings belonging to the spanning cluster and the total number of strings in the system.

\subsection{Density profiles}

\begin{figure}[ht]
\centering
\includegraphics[scale=0.6]{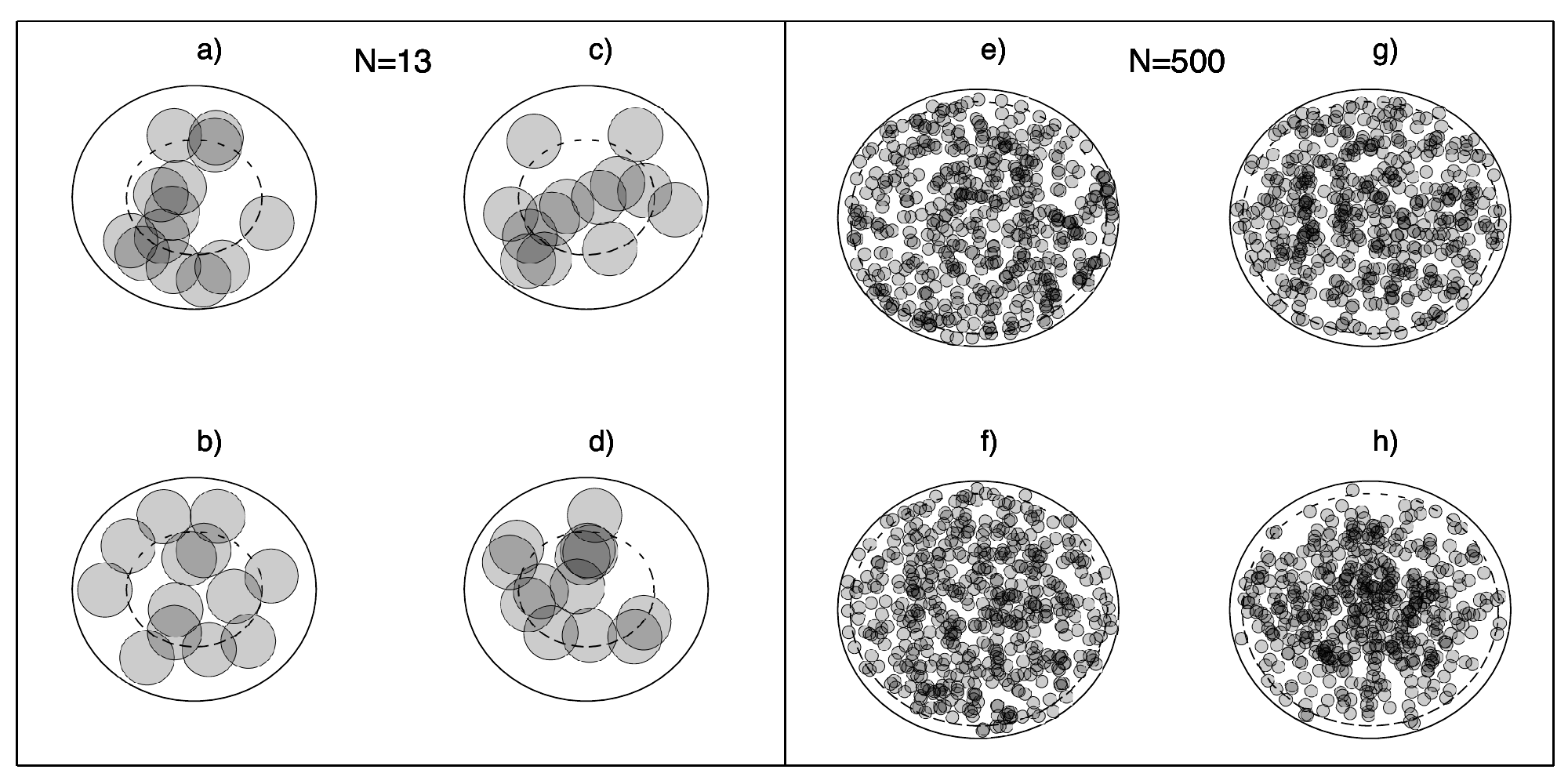}
\caption{Samples of a percolating system for different density profiles. Left box: String systems at $N=13$, $\eta=0.7$ and $\varepsilon=0.4$, for the models a) Uniform, b) $1s$, c) $2^{1/2}s$ and d) $2s$;  Right box: String systems at $N=500$, $\eta=1.1$ and $\varepsilon=0.4$, for the models e) Uniform, f) $1s$, g) $2^{1/2}s$ and h) $2s$.}
\label{sample13500}
\end{figure}

Commonly the continuum percolation theory studies Uniform distributions of objects. Nevertheless, in heavy-ion collisions the nuclear profile function is considered in a more realistic way, denser in the central region of the nucleus and being more dilute as we go away from the central region.

\begin{figure}[ht]
\centering
\includegraphics[scale=1]{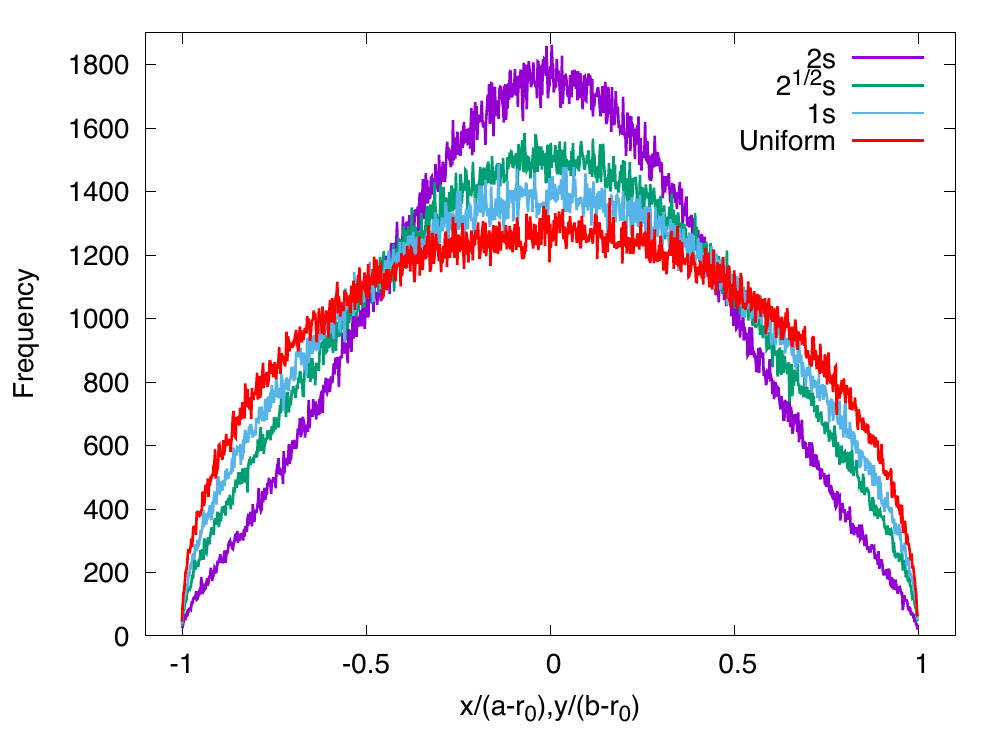}
\caption{Projection of the disc position distribution over the axis of the ellipse for the different profile functions. The histograms were built with a $10^6$ generated positions for each model. The distributions are normalized to the semi-axis value to eliminate the dependence on the number of discs $N$, $\varepsilon$ and $\eta$.}

\label{dots2}
\end{figure}
To consider different density profiles in pp collisions, we consider three Gaussian distributions functions, where the second distribution corresponds to the one taken in Ref. \cite{Rodrigues:1998it} and the other two within two extremes of Gaussian profiles. 
For the Gaussian density profiles we generate the position discs $(x,y)$ with the following probability distribution function:

\begin{equation}
f(x,y)=\frac{1}{2\pi \sigma_a\sigma_b}\exp\left[ -\frac{1}{2}\left(\frac{x^2}{\sigma_a^2}+\frac{y^2}{\sigma_b^2} \right) \right],
\end{equation}
where $\sigma_a$ and $\sigma_b$ are standard deviations over the semi-axis of the ellipse. In the following we will consider the cases:

\begin{subequations}
\begin{eqnarray}
\sigma_a=(a-r_0),&\sigma_b=(b-r_0); \label{1s}\\
\sigma_a=(a-r_0)/2^{1/2},&\sigma_b=(b-r_0)/2^{1/2}; \label{sq}\\
\sigma_a=(a-r_0)/2,&\sigma_b=(b-r_0)/2. \label{2s}
\end{eqnarray}
\end{subequations}
The models with different standard deviations are denoted by the denominator as $1s$, $2^{1/2}s$ y $2s$ corresponding to the equations ~\eqref{1s},~\eqref{sq} and ~\eqref{2s}, respectively.
Note that the denoted $2^{1/2}s$ model for $\varepsilon=0.00$ reproduces the density profile shown in Ref. \cite{Rodrigues:1998it}.

The percolating systems examples with eccentricity $\varepsilon=0.40$ for the different density model profiles for $N=13$ (left box) and $N=500$ (right box) are presented in Fig. \ref{sample13500}. The disc position projection distribution over the axis of the ellipse for the four different profile functions are plotted on Fig.~\ref{dots2}, where disc position projection over each ellipse axis has been normalized according to the semi-axis value, leading to an independent distribution of the eccentricity, the string density and the number of clusters of the system. Note that the Gaussian density profiles increase the string or particle concentration in the center of the ellipse, as exhibited in Fig. ~\ref{dots2}, where the disc position projections over each semi-axis of the ellipse are shown.

The Uniform, $1s$ and $2^{1/2}s$ models do not show significant differences in the disc distributions for a small number of strings, but we expect to have larger differences as we increase the number of strings Ref. \cite{Rodrigues:1998it}. As the number of peripheral discs decrease the percolation threshold becomes higher, such as in the $2^{1/2}s$ model, where $\eta_c \sim 1.59$ was reported in Ref. \cite{Rodrigues:1998it}.

\section{Results}

To determine the behavior of the percolation probability as a function of filling factor and eccentricity for the number of strings in the system: $N=$13, 55, 96, we start from an initial filling factor value $\eta=0.10$, with step increments of the size $\Delta \eta= 0.05$ until the value of $\eta=1.8$ is reached. 
To obtain $P$ as a function of $\varepsilon$, we begin with the limit case $\varepsilon=0$, which is the classical model of string percolating system considering a circle boundary. In the simulation, $2\times10^{5}$ realizations were performed for each filling factor and eccentricity values to all numbers of strings given, for each density profile model. The results presented in this section will only consider the Uniform distribution model but similar results are obtained for the Gaussian distribution models.

\begin{figure}[H]
\includegraphics[scale=0.75]{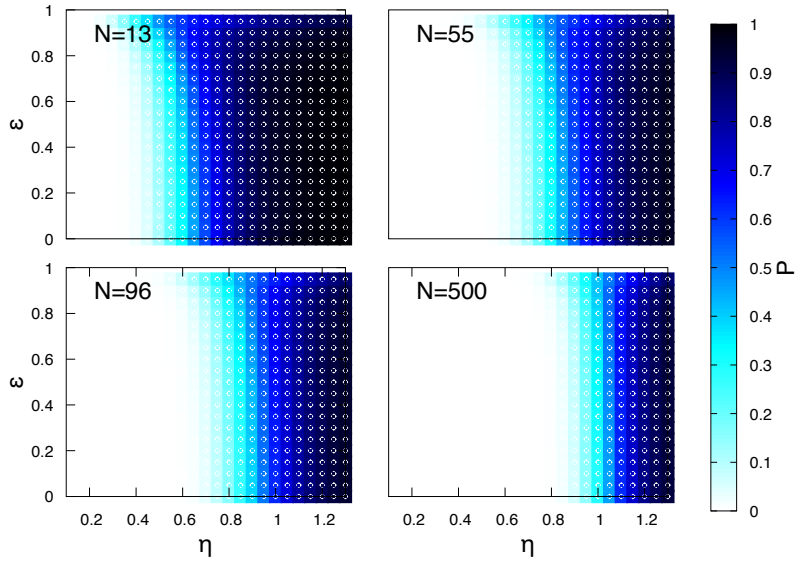}
\caption{Percolation probability $P$ as a function of filling factor $\eta$ and eccentricity $\varepsilon$ for different values of a number of strings $N$ with Uniform density profile. }
\label{fig-p-varios}
\end{figure}

The behavior of $P$ as a function of $\eta$ and $\varepsilon$ for $N=13,55,96,500$, where the $x-$axis is the filling factor and the $y-$axis is the eccentricity, with the Uniform model is shown in Fig.~\ref{fig-p-varios}. The value of the percolation probability as a function of $\eta$ and $\varepsilon$ is shown in gradient colors, where light and dark colors correspond to the values close to zero and the unit respectively.

Blue tones on Fig.~\ref{fig-p-varios} show the region where the percolation transition appears in the Uniform model. The percolation probability shows an evident shift as a function of $N$ and high eccentricity dependence associated with a finite size effect. 

\begin{figure}[H]
\centering
\includegraphics[scale=.4]{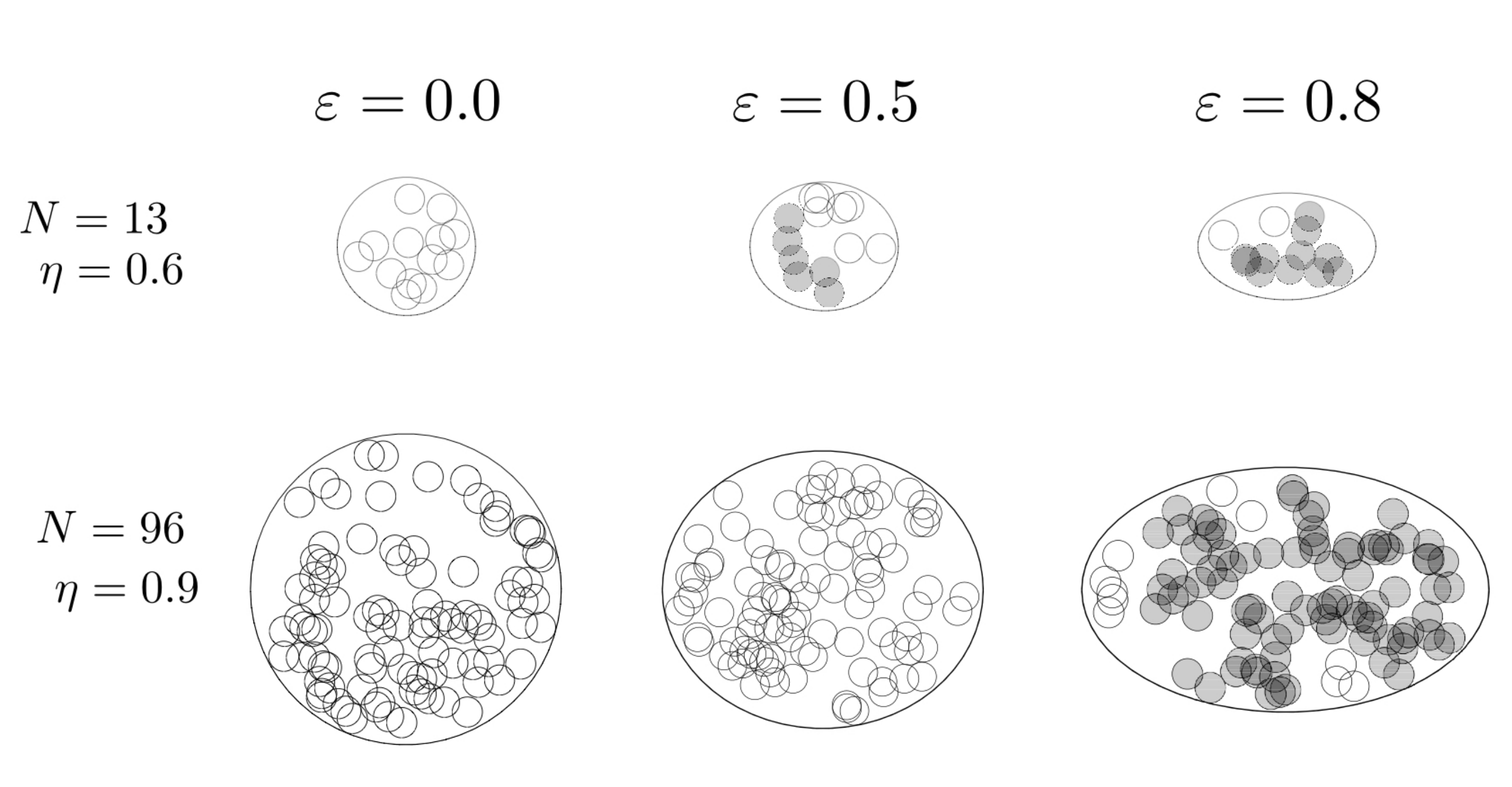}
\caption{Spanning cluster as a function of the eccentricity for $N=$13, 96 and different filling factor in Uniform model. Filled circles are strings belonging to the spanning cluster.}
\label{fig-perife}
\end{figure}

In Fig.~\ref{fig-perife} we show the emergence of the spanning cluster as a function of the eccentricity for small values of $N$ in the Uniform model. The spanning cluster emerges when we increase the values of the filling factor as we increase values of $N$. For largest values of $N$, the percolation probability becomes independent of the eccentricity and the phase transition appears around the percolation threshold for the continuum percolation in the thermodynamic limit. 

\begin{figure}[H]
\includegraphics[scale=0.75]{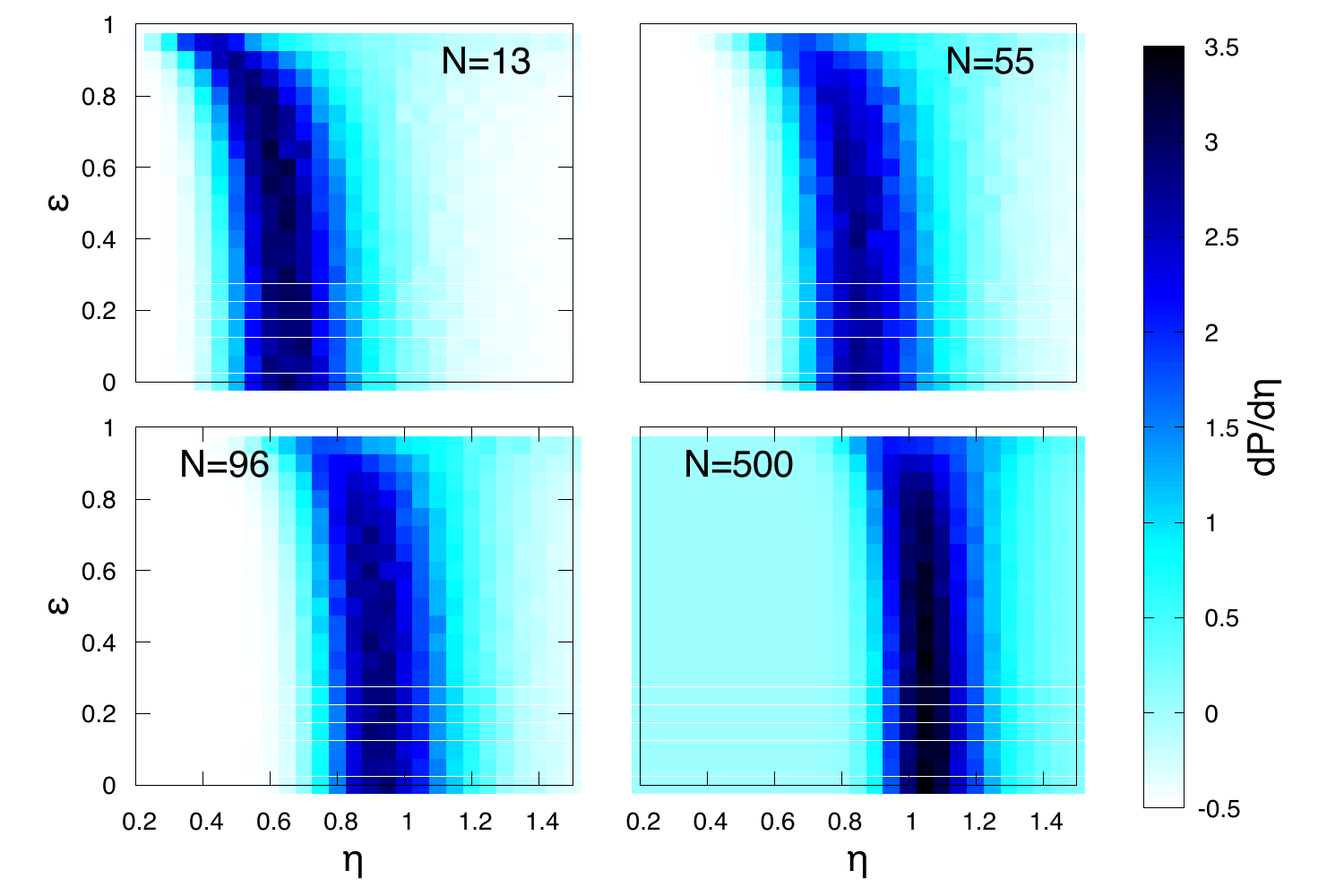}
\caption{A derivative of the percolation probability with respect to filling factor $\eta$ in the Uniform model.}
\label{fig-dp-varios}
\end{figure}

The percolation threshold for a percolating system can be  quantified from the percolation probability. An approximate allowed region where the percolation threshold appears when the derivative $\partial P /\partial \eta$ takes its maximum value. Results of the numerical derivative $\partial P/ \partial \eta$ are shown in Fig.~\ref{fig-dp-varios}. Dark tones correspond to the region where $\partial P/ \partial \eta$ takes its maximum value.

\begin{figure}[H]
\centering
\includegraphics[scale=0.7]{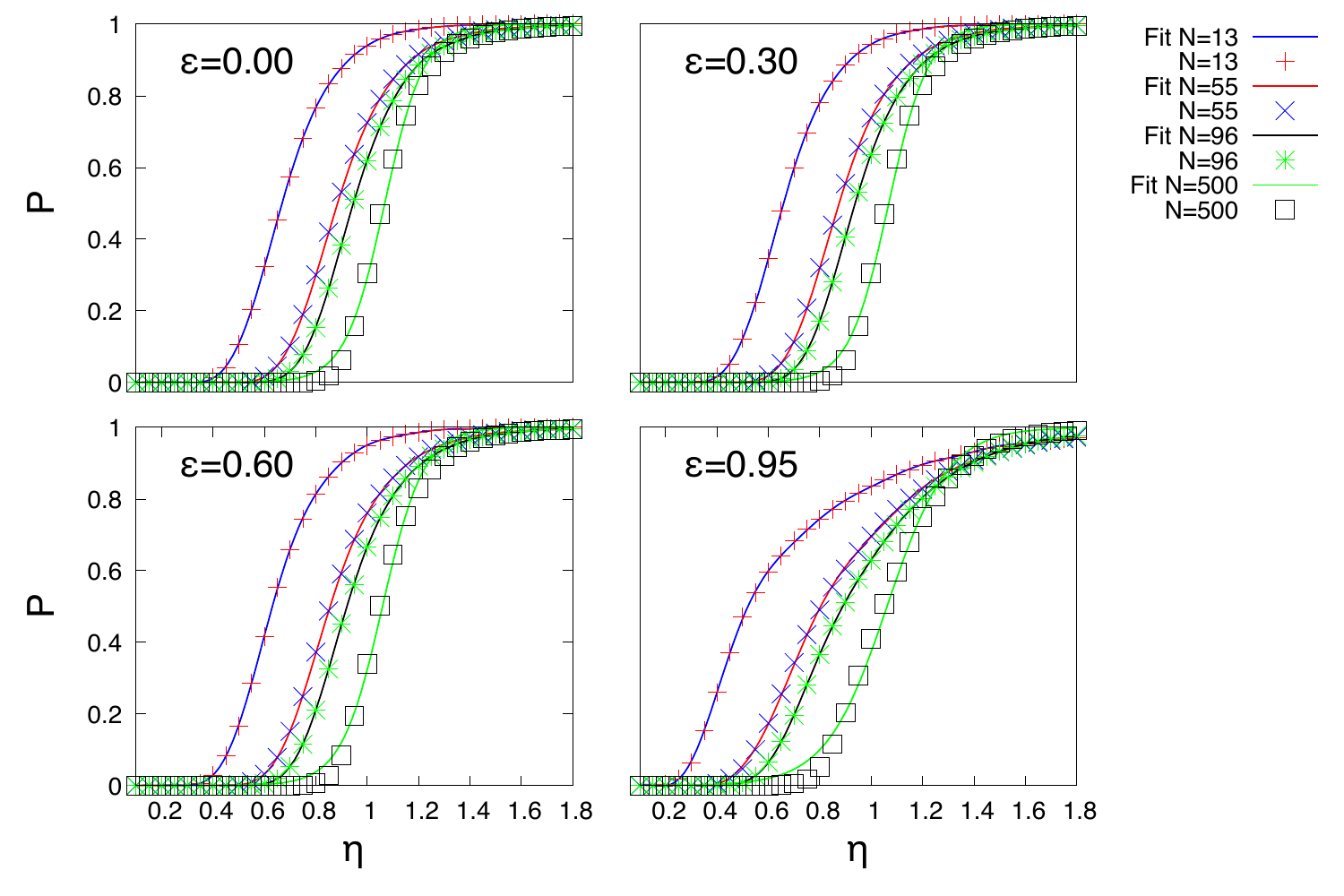}
\caption{Optimum fit for the percolation probability according to Eq.~\eqref{fit1} to $N=13,55,96$ and Eq.~\eqref{fit2} to $N=500$ for the given eccentricities $\varepsilon=0.00,0.30,0.60,0.95$. }
\label{fig-fit}
\end{figure}

\begin{figure}[H]
\centering
\includegraphics[scale=0.9]{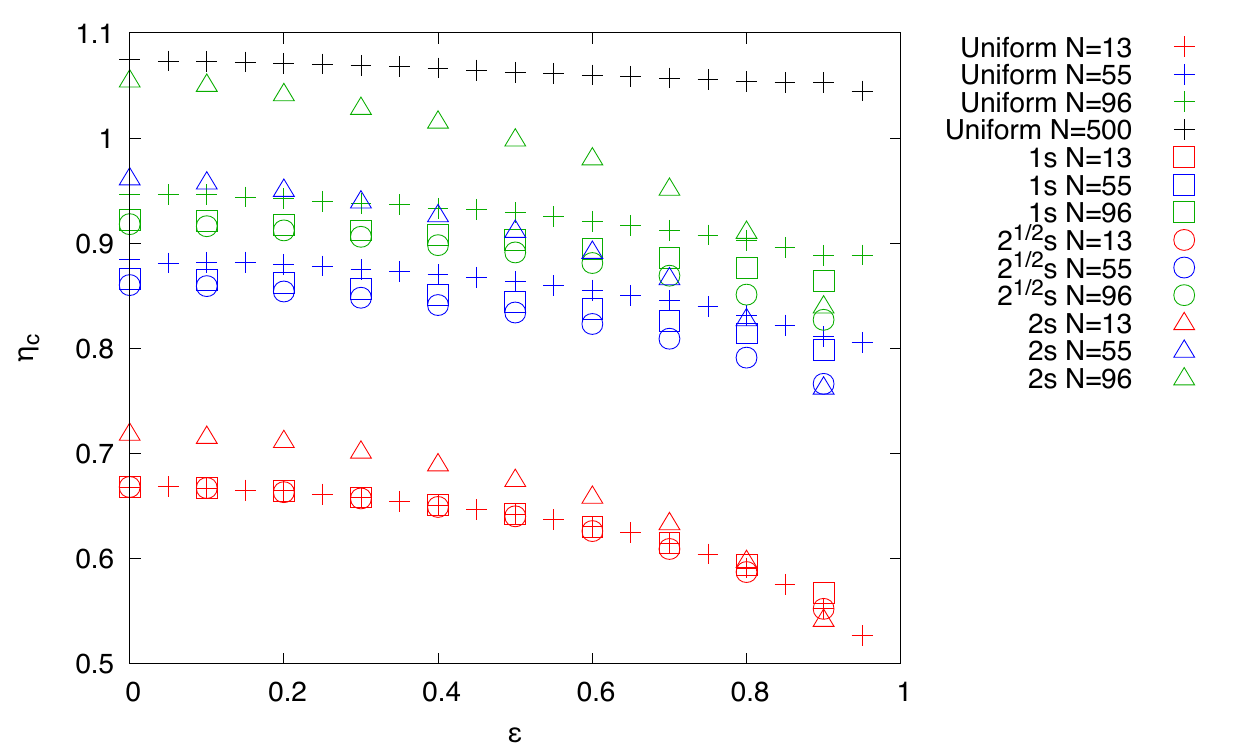}
\caption{Percolation threshold $\eta_c$ as a function of eccentricity ($\varepsilon$) for different values of $N$ with the different density profiles: Uniform (crosses), and Gaussian: $1s$ (squares), $2^{1/2}s$ (circles) and $2s$ (triangles). }
\label{fig-pc}
\end{figure}

For a precise determination of the percolation threshold $\eta_c$, according to the method in Ref.~\cite{1997JPhA...30L.585R,Li}, we fit the percolation probability for each value of $\varepsilon$ with a function of the form
\begin{equation}
P(\eta)=\frac{1}{1+\exp\left(-\sum_{k=0}^4a_k\eta^k\right)}.
\label{fit1}
\end{equation}
To obtain the value of $\eta_c$, the equation $P(\eta_c)=0.5$ has to be solved \cite{1997JPhA...30L.585R}. However, for $N=500$ and the model with Gaussian density profiles, we use
\begin{equation}
P(\eta)=\frac{1}{2}\left[ 1+\tanh\left(  \frac{\eta-\eta_c}{\Delta L} \right)  \right],
\label{fit2}
\end{equation}
where $\eta_c$ is the percolation threshold and $\Delta L$ is the width of the percolation transition \cite{1997JPhA...30L.585R}.
In Fig.~\ref{fig-fit}, we show the best fit obtained for the percolation probability as a function of filling factor for different values of $\eta$ and $N$ in string percolating systems with Uniform density profile.  
The behavior of the percolation threshold as a function of $\eta$ for systems with $N=13,55,96$ strings with the corresponding profile densities: Uniform (crosses), $1s$ (squares), $2^{1/2}s$ (circles) and $2s$ (triangles) is exhibit in Fig.\ref{fig-pc}. The case for $N=500$ with the Uniform model is also shown as an approach to the thermodynamical limit.

\section{Discussion and conclusions}

We have extended the String Percolation Model to study the behavior of the percolation threshold
with the initial state geometry description for the continuum percolation model with a finite number of discs. The dependence of this threshold is described with the filling factor and the eccentricity of the elliptical boundary system. 

When a boundary without periodic conditions is impose, the system presents a finite size effect which is very notable for a system with a small number of strings ($N\sim 13$). In particular, the percolation threshold presents a strong dependence on the eccentricity for a small $N$. This is originated due to the approach of the cross-over length magnitude to the string diameter at eccentricities greater than $0.7$. We observe that the percolation threshold for elliptical bounded systems is shifted to smaller values with respect to the circular boundary case. Complementary, for high eccentricities the system needs shorter distances to cross-over.

Furthermore, for small systems, there are no significant differences for the percolation threshold between the Uniform, the $1s$ and $2^{1/2}$ models due to the fact the spatial distribution of the disc inside the ellipse is very similar as shown in Fig. ~\ref{dots2}.

On the other hand, for highly populated systems (large values of $N$) the system becomes independent of the eccentricity because this condition corresponds to the thermodynamic limit, where there is no difference between large circles and ellipses compared with one string area. Moreover, for the string percolation systems with Uniform density profile, the percolation threshold becomes closer to the value $\eta_c\sim 1.13$ as it is reported for continuum percolation in the thermodynamic limit \cite{Mertens,Satz:2002ku,Digal:2003sg,Ferreiro:1998jr,Isichenko:1992zz,Quintanilla}. Besides in the Gaussian profile models, the spanning cluster is formed by a bigger number of strings than in the Uniform model under the same filling factor and eccentricity conditions. 

Note that as the standard deviation in the Gaussian models decreases the string distribution will differ more from the Uniform density profile. In particular, for small systems, we can assume that if we consider the $2^{1/2}s$ model as a limit case where the dispersion is such that $\sigma_a /a >1/\sqrt{2}$ and $\sigma_b /b >1/\sqrt{2}$, the density profiles will have percolation thresholds comparable to the Uniform case.

To conclude we have shown that the finite number of initial strings and the boundary initial state geometry effects will become relevant in the description of the SPM of pp collisions systems since the number of strings involved is much more small compare to the ones in nuclear collisions.
Implications of the role of considering an elliptical boundary on the initial state geometry for small number of strings systems will give a correction factor to the usual parameters of the model which relevance for physical observables of Quark Gluon Plasma formation in pp collision systems is under discussion and will be reported in a future work. 

\section{Acknowledgments}
This work has been supported by CONACyT. The computer simulations were performed at the LNS-BUAP.

\newpage

\section{References}

 \bibliographystyle{elsarticle-num} 
 \bibliography{SPMebs.bbl}

\begin{thebibliography}{10}
\expandafter\ifx\csname url\endcsname\relax
  \def\url#1{\texttt{#1}}\fi
\expandafter\ifx\csname urlprefix\endcsname\relax\def\urlprefix{URL }\fi
\expandafter\ifx\csname href\endcsname\relax
  \def\href#1#2{#2} \def\path#1{#1}\fi

\bibitem{Broadbent:1957rm}
S.~R. Broadbent, J.~M. Hammersley, Proc. Cambridge Phil. Soc. 53 (1957)
  629--641.

\bibitem{Hoshen:1976zz}
J.~Hoshen, R.~Kopelman, Phys. Rev. B14 (1976) 3438--3445.

\bibitem{schulman1986}
L.~S. Schulman, P.~E. Seiden, Science 233~(4762) (1986) 425--431.

\bibitem{Klypin:1993rz}
A.~Klypin, S.~F. Shandarin, The Astrophysical Journal 413 (1993) 48--58.

\bibitem{Biro:1986ebt}
T.~S. Biro, J.~Knoll, J.~Richert, Nucl. Phys. A459 (1986) 692--710.

\bibitem{Baym}
G.~Baym, Physica A: Statistical Mechanics and its Applications 96~(1) (1979)
  131 -- 135.

\bibitem{Campi:1986nxv}
X.~Campi, J. Phys. Colloq. 47~(C4) (1986) 419--422.

\bibitem{Kirkpatrick}
S.~Kirkpatrick, Rev. Mod. Phys. 45 (1973) 574--588.

\bibitem{Stauffer:1978kr}
D.~Stauffer, Phys. Rept. 54 (1979) 1--74.

\bibitem{Elliott:2002dk}
J.~B. Elliott, et~al., Phys. Rev. C67 (2003) 024609.

\bibitem{Stauffer-book}
D.~Stauffer, A.~Aharony, Introduction To Percolation Theory, Taylor \& Francis,
  1994.

\bibitem{Saberi}
A.~A. Saberi, Physics Reports 578 (2015) 1 -- 32.

\bibitem{feng}
D.~Feng, G.~Jin, Introduction to Condensed Matter Physics, no. v. 1, World
  Scientific, 2005.

\bibitem{sykes}
M.~F. Sykes, D.~S. Gaunt, M.~Glen, Journal of Physics A: Mathematical and
  General 9~(5) (1976) 725.

\bibitem{Essam}
J.~W. Essam, Reports on Progress in Physics 43~(7) (1980) 833.

\bibitem{Rafelski:2013obw}
J.~Rafelski[Nucl. Phys. Proc. Suppl.243-244,155(2013)].

\bibitem{Bala:2016hlf}
R.~Bala, I.~Bautista, J.~Bielcikova, A.~Ortiz, Int. J. Mod. Phys. E25~(07)
  (2016) 1642006.

\bibitem{Pasechnik:2016wkt}
R.~Pasechnik, M.~Sumbera, Universe 3~(1) (2017) 7.

\bibitem{Armesto:2015ioy}
N.~Armesto, E.~Scomparin, Eur. Phys. J. Plus 131~(3) (2016) 52.

\bibitem{Foka:2016vta}
P.~Foka, M.~A. Janik, Rev. Phys. 1 (2016) 154--171.

\bibitem{Braun:2015eoa}
M.~A. Braun, J.~Dias~de Deus, A.~S. Hirsch, C.~Pajares, R.~P. Scharenberg,
  B.~K. Srivastava, Phys. Rept. 599 (2015) 1--50.

\bibitem{Bali:1994de}
G.~S. Bali, K.~Schilling, C.~Schlichter, Phys. Rev. D51 (1995) 5165--5198.

\bibitem{Schlichter:1994dc}
C.~Schlichter, G.~S. Bali, K.~Schilling, Nucl. Phys. Proc. Suppl. 42 (1995)
  273--275.

\bibitem{Nardi:1998qb}
M.~Nardi, H.~Satz, Phys. Lett. B442 (1998) 14--19.

\bibitem{Amelin:1993cs}
N.~S. Amelin, M.~A. Braun, C.~Pajares, Phys. Lett. B306 (1993) 312--318.

\bibitem{Braun:2001us}
M.~A. Braun, F.~Del~Moral, C.~Pajares, Phys. Rev. C65 (2002) 024907.

\bibitem{Celik:1980td}
T.~Celik, F.~Karsch, H.~Satz, Phys. Lett. B97 (1980) 128--130.

\bibitem{Bautista:2015kwa}
I.~Bautista, A.~Fernandez~T\'ellez, P.~Ghosh, Phys. Rev. D92 (2015) 071504.

\bibitem{I.Bautista:2017}
R.~Alvarado, I.~Bautista, P.~Fierro, PoS (ICHEP2016) 1152.

\bibitem{Schenke:2014jza}
B.~Schenke, P.~Tribedy, R.~Venugopalan, Nucl. Phys. A931 (2014) 288--292.

\bibitem{Gelis:2016upa}
F.~Gelis, B.~Schenke, Ann. Rev. Nucl. Part. Sci. 66 (2016) 73--94.

\bibitem{Bautista:2009my}
I.~Bautista, L.~Cunqueiro, J.~D. de~Deus, C.~Pajares, J. Phys. G37 (2010)
  015103.

\bibitem{Rodrigues:1998it}
A.~Rodrigues, R.~Ugoccioni, J.~Dias~de Deus, Phys. Lett. B458 (1999) 402--406.

\bibitem{Mertens}
S.~Mertens, C.~Moore, Physical Review E - Statistical, Nonlinear, and Soft
  Matter Physics 86~(6) (2012) 1--6.

\bibitem{Reynolds:1980zz}
P.~J. Reynolds, H.~E. Stanley, W.~Klein, Phys. Rev. B21 (1980) 1223--1245.

\bibitem{Ziff}
R.~M. Ziff, B.~Sapoval, Journal of Physics A: Mathematical and General 19~(18)
  (1986) L1169.

\bibitem{Newman}
M.~E.~J. Newman, R.~M. Ziff, Phys. Rev. Lett. 85 (2000) 4104--4107.

\bibitem{wexler}
C.~Wexler, Addison-wesley series in introductory mathematics, Addison-Wesley
  Publishing Company, 1961.

\bibitem{Satz:2002ku}
H.~Satz, Nucl. Phys. A715 (2003) 3--19.

\bibitem{1997JPhA...30L.585R}
M.~D. {Rintoul}, S.~{Torquato}, Journal of Physics A Mathematical General 30
  (1997) L585--L592.

\bibitem{Li}
C.~Li, T.~W. Chou, Applied Physics Letters 90 17.

\bibitem{Digal:2003sg}
S.~Digal, S.~Fortunato, H.~Satz, Eur. Phys. J. C32 (2004) 547--553.

\bibitem{Ferreiro:1998jr}
E.~G. Ferreiro, C.~Pajares, Nucl. Phys. A642 (1998) 143--148.

\bibitem{Isichenko:1992zz}
M.~B. Isichenko, Rev. Mod. Phys. 64 (1992) 961--1043.

\bibitem{Quintanilla}
J.~A. Quintanilla, R.~M. Ziff, Phys. Rev. E 76 (2007) 051115.

\end{thebibliography}

\end{document}